\documentclass[useAMS,usenatbib]{mn2e}
\usepackage{graphicx}
\usepackage{epsfig}

\begin{document}
\title{Evidence on the Origin of Ergospheric Disk Field Line Topology in Simulations of Black Hole Accretion}
\author[Brian Punsly] {Brian Punsly \\ 4014 Emerald Street No.116, Torrance CA, USA 90503 \\
ICRANet, Piazza della Repubblica 10 Pescara 65100, Italy,\\
E-mail: brian.punsly@verizon.net or brian.punsly@comdev-usa.com\\}
\maketitle \label{firstpage}
\begin{abstract}
This Letter investigates the origin of the asymmetric magnetic field line
geometry in the ergospheric disk (and the corresponding asymmetric powerful jet) in 3-D perfect magnetohydrodynamic
(MHD) numerical simulations of a rapidly rotating black hole
accretion system reported in \citet{pun10}. Understanding,
why and how these unexpected asymmetric structures form is of practical interest
because an ergospheric disk jet can boost the black hole driven jet power
many-fold possibly resolving a fundamental disconnect between
the energy flux estimates of powerful quasar jets and simulated jet power \citep{pun11}.
The new 3-D simulations of \citet{bec09} that were run with basically the same code that was used in
the simulation discussed in \citet{pun10} describe the "coronal
mechanism" of accreting poliodal magnetic flux towards the event horizon.
It was determined that reconnection in the inner accretion disk
is a "necessary" component for this process. The coronal mechanism seems
to naturally explain the asymmetric
ergospheric disk field lines that were seen in the simulations.
Using examples from the literature, it is discussed how apparently small changes in the
reconnection geometry and rates can make enormous changes in the
magnetospheric flux distribution and the resultant black hole driven
jet power in a numerical simulation. Unfortunately,
reconnection is a consequence of numerical diffusion and not a detailed
(yet to be fully understood) physical mechanism in the existing
suite of perfect MHD based numerical simulations. The implication is
that there is presently great uncertainty in the flux distribution
of astrophysical black hole magnetospheres and the resultant jet
power.
\end{abstract}

The ultimate nature of the power output of a black hole
magnetosphere is highly dependent on two poorly understood
circumstances, the source of plasma injection on the magnetic field
lines that thread the event horizon and the fate of accreted
vertical magnetic flux \citep{pun90,pun91}. In the past eight years,
the use of perfect magnetohydrodynamic (MHD) numerical simulations
have been developed in the scientific community to help understand
these issues. In order to establish and maintain the event horizon
magnetosphere, the simulations must rely on the numerical artifice
of a mass floor, local mass injection to establish a minimum
density. In spite of some discussion of anecdotal mass floor
examples, perfect MHD simulations are not likely to resolve the
first issue since a mass floor violates rest-mass and
energy - momentum conservation and necessarily contradicts the perfect
MHD assumption \citep{mck06}. However, recently MHD simulations have
shed some light on the second point. The mystery of how magnetic
flux was accreted to the black hole in MHD numerical simulations was
revealed in \citet{bec09} through their meticulous high resolution
numerical work. In this Letter, the insight provided by
\citet{bec09} is used to explain the strange ergospheric disk field
topology seen in the simulations reported in \citet{pun10} that
resulted in powerful one-sided jets that dominate the total energy
output in the jetted system with a location that changed hemispheres
in different time snapshots \citep{pun07,pun08}.
\par The fate of accreted flux is perhaps the most critical issue in understanding the power source for
black hole driven jets. It was noted in \citet{pun91}, that if
vertical, magnetic flux accretes, it is not clear where it ends up.
It was argued that reconnection of vertical flux would be
determinant to the final magnetic field configuration since the
black hole is effectively a sink with infinite capacity for mass,
but with a very limited capacity to accept magnetic flux. One
possible field configuration that could result from reconnection
produced a disk in the ergosphere that can drive powerful jets. If
this happens, the power source for the jet drastically increases in
efficiency, so this configuration is of profound interest in AGN
\citep{nem07,pun11}. However, there is much scientific uncertainty
in the reconnection geometry and rate expected in an accretion flow
near a rapidly rotating black hole, since our experimental
experience is based on very different environments, the solar
corona, the solar wind, the Earth magnetosphere and magneto-tail,
and magnetic confinement devices for thermonuclear fusion. The
situation was further complicated by the argument that the existence
of coherent vertical flux within the dense accreting gas is
inhibited by the turbulent magnetic diffusivity of the plasma
\citep{van89,lub94}. Thus, it was not even clear if the notion of a
large scale field associated with an accretion flow was viable.
\par However, recent numerical simulations have shown that
vertical flux accretion can occur, but not by diffusing through the
disk, but by a two step process called the "coronal mechanism". The
first step is the transport towards the black hole of an inwardly
stretched, disk-anchored, poloidal loop through the low turbulence,
coronal layer just above the disk, the "hairpin" field in
\citet{bec09}. This coronal transport is similar to the mechanism
proposed in \citet{rot08}, but see \citet{bec09} for some dynamical
differences. Then loops of magnetic field in the inner accretion
flow reconnect with the half of the "hairpin" at mid-latitudes,
allowing it to contract into the black hole, leading to one sign of
field threading the black hole (see section 2). Reconnection is
apparently "necessary" for vertical flux accretion in all the
numerical simulations of this family \citet{bec09,haw06,mck09}, but
the geometry of the reconnection site is very different than what
was described in \citet{pun91}. In spite of this geometric
difference, an ergospheric disk does form in some simulations. In
the following, the coronal mechanism is shown to naturally explain
the asymmetric magnetic field observed in 3-D simulations of the
ergospheric disks \citep{pun10}.
\par This work derives from a family of simulations based on a constrained
transport MHD code on the Kerr spacetime background that has been described numerous times in the literature,
so this is not reproduced here \citep{dev02,dev03,dev05,hir04,kro05,haw06,bec08}. The particular simulation,
KDJ, discussed in \citet{pun10} was described in detail in
\citet{kro05,haw06}, which the reader should consult for particulars. KDJ simulates
accretion onto a rapidly rotating black hole with an angular momentum per unit mass of a = 0.99 M
in geometrized units. This Letter focuses on the very complicated twisted
topologies that are involved in the reconnection events in the simulation.
\section{Field Line Topology in the Ergospheric Disk}
A surprising feature of 3-D MHD numerical simulations is that
the ergospheric disk was threaded mainly by the type I field lines
instead of type III field lines in the nomenclature introduced in \citet{pun10}. In the left hand
panel of Figure 1, the type I vertical magnetic field lines emerge
from the inner equatorial accretion flow. The false color plot is a
contour map of a 2-D cross-section of the density in Boyer-Lindquist
coordinates expressed in code units. The interior of the inner
calculation boundary (just outside of the event horizon) is
grayish-white. These field lines are distinguished by connecting to
the Poynting jet in one hemisphere only, with the other end
spiraling around within the accreting gas in the opposite
hemisphere. Another distinguishing feature of a type I field line
is that the azimuthal direction of the magnetic field changes
direction as the field line crosses the midplane of the accretion
flow. It was expected in \citet{pun91} that the direct advection of
magnetic flux would produce the type III field topology depicted in
the right hand frame of Figure 1. But these were rare in the 3-D
simulation reported in \citet{pun10}. Unlike the type I field
lines, the type III field lines connect to both sides of the bipolar
jet.
\par One should consult Figure 3 of \citet{pun10} to appreciate the physical significance
of the one sided type I field lines from the ergopsheric disk. These
field lines directly equate to a powerful jet of Poynting flux. When
this jet forms, it swamps the power output from the event horizon
\citep{pun07,pun08}. In fact, it even suppresses the event horizon
jet power to some degree \citep{pun11}.
\begin{figure*}
\includegraphics[scale=0.11]{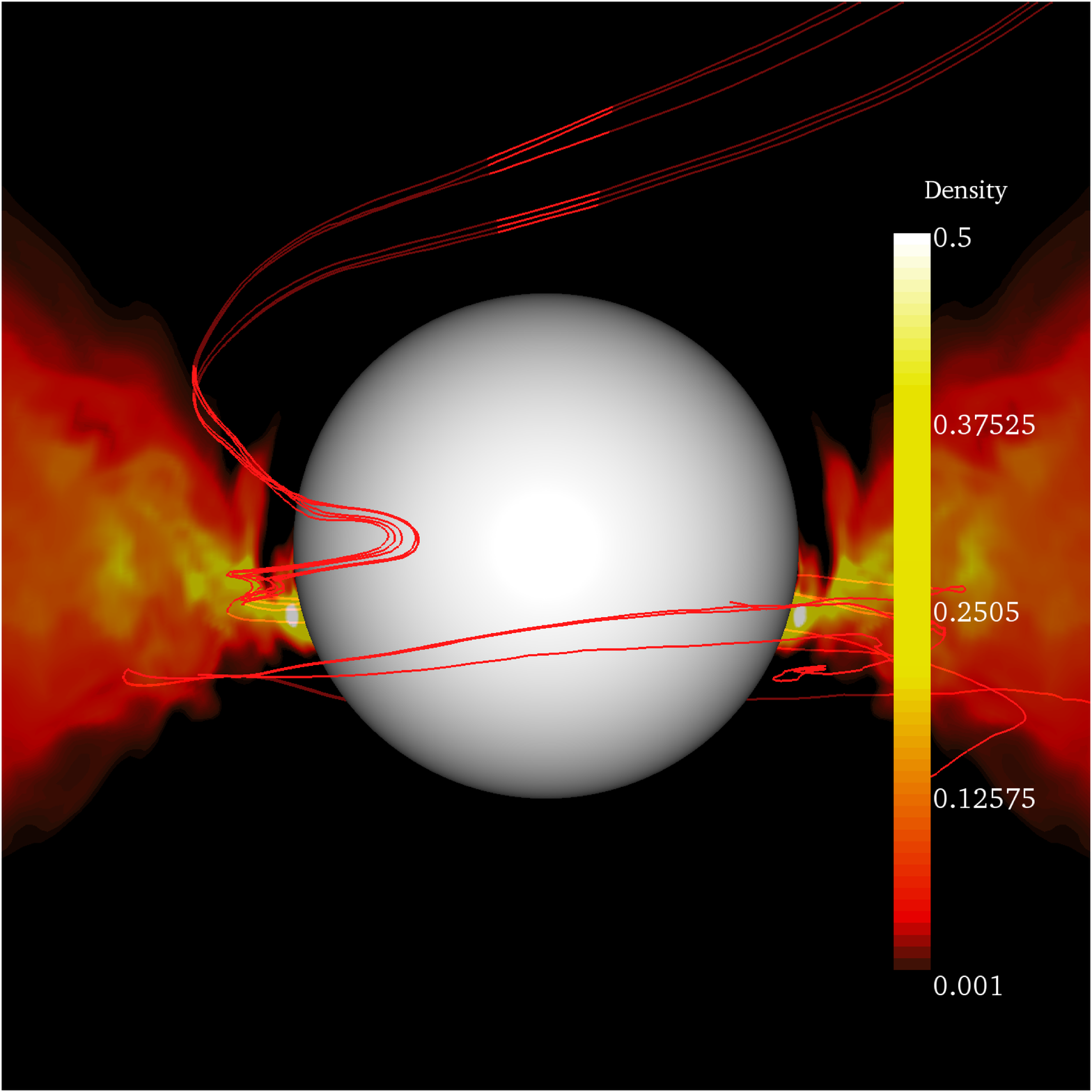}
\includegraphics[scale=0.11]{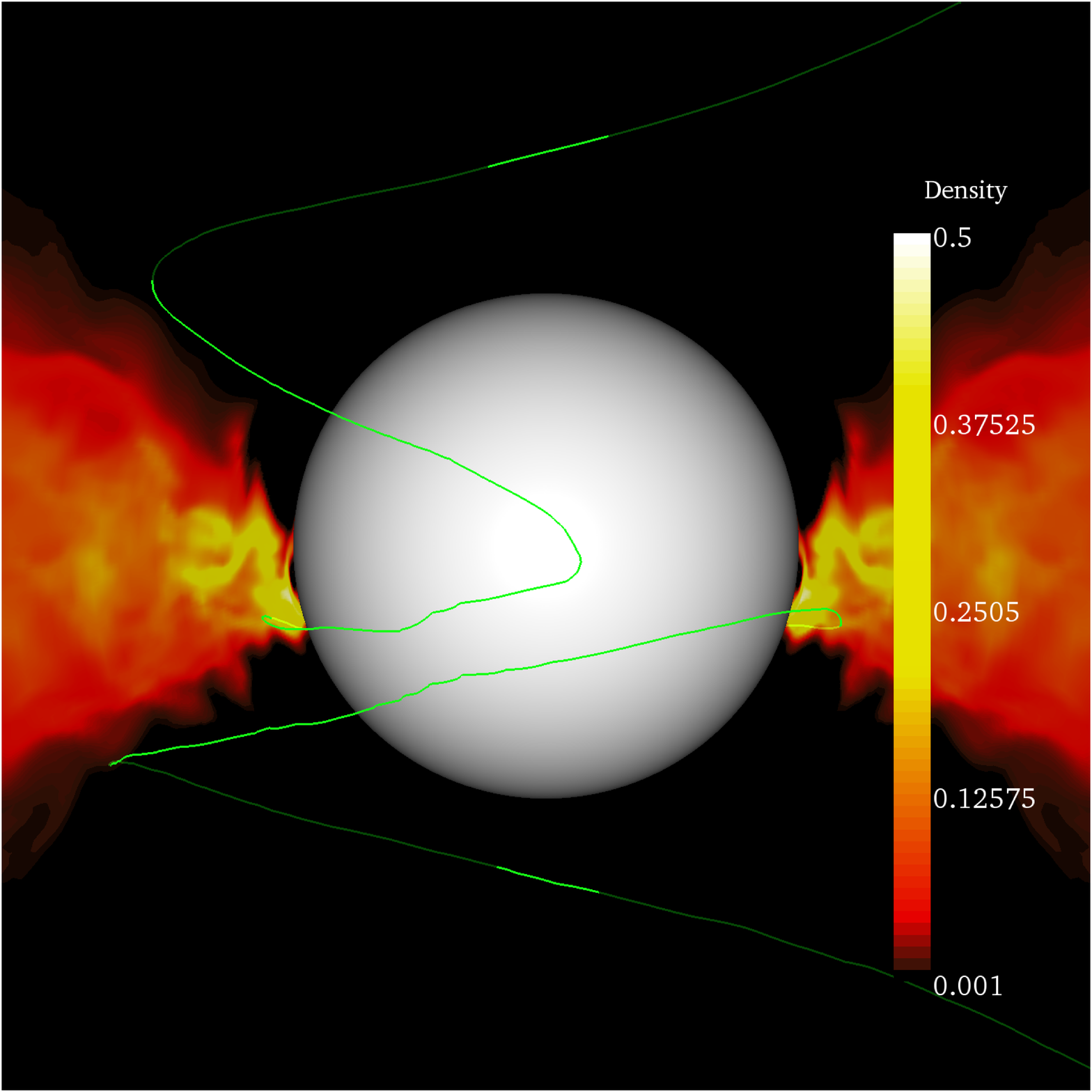}
\caption{Plots of ergospheric disk field lines in the simulation,
KDJ, from Punsly et al (2009). For details of how the field lines
were plotted and the false color 2-D density cross section, please
refer to that paper. It was expected that the type III ergospheric
disk field line topology (on the right) would dominate the vertical
flux through the ergospheric disk. Instead 3-D MHD numerical
simulations found very few type III field lines, but many type I
field lines (on the left). It should be noted that the images are
3-D. They were created in the 3-D visualization tool, Paraview 3.3,
as 3-D images and the "camera" is placed in the equatorial plane
(i.e., if the y-axis points toward the camera, the gas density is
plotted in the x-z plane). The field lines that pass behind the x-z
plane are made darker (i.e. the x-z plane is partially opaque, 75\%)
to make it easy to distinguish them from the magnetic coils in front
of the x-z plane. However, near the poles they are not darker
because a conical region of the numerical grid is excised in that
region (therefore, there are no points in the x-z plane above the
poles of the horizon and no opacity) as a numerical expedience as
discussed in Hawley and Krolik (2006) and Punsly et al (2009).}
\end{figure*}
\section{Relevant Aspects of the Coronal Mechanism} In this section, the coronal mechanism is reviewed
and asymmetric variations in the process are noted that are relevant
to the following. The coronal mechanism was detailed in Figure 11 of
\citet{bec09}. Some important geometric simplifications were
implemented in \citet{bec09} that greatly improved the clarity of
presentation. In that paper, the primary focus was on detailed
discussions of symmetric pairs of "hairpin" field lines, where one
approaches from just above the accretion disk from the North and a
matching "hairpin" field line approaches from the South.
Furthermore, a non-rotating black hole was chosen so there is
minimal azimuthal twisting near the event horizon. Ostensibly, for
the sake of simplicity of presentation, all the 3-D data was
averaged over azimuth in \citet{bec09}, so there were simple field
line topologies like closed 2-D poloidal loops in the accretion disk
(see their Figure 11).
\par After the leading edge of the hairpins have
penetrated the event horizon, reconnection occurs in the equatorial
portion of the hairpin with a closed 2-D loop. This leads to
an increase in the magnetic flux in the accretion flow vortex, or "funnel", in the
polar region beyond the event horizon radius in both hemispheres.
\par It is noted here that there are two intriguing aspects of the field line evolution
presented in \citet{bec09} that should be relevant to 3-D field line
reconnection in the Kerr geometry.
\begin{enumerate}
\item There is
not always symmetry in the hairpin accretion between the northern
and southern hemispheres. A cluster of field lines in a "hairpin"
configuration often approach from one hemisphere at a time. This is
manifested in the on-line animation for \citet{bec09}, especially
around 23 seconds, where one sees an excess of organized flux in the
funnel, in the southern hemisphere of the event horizon. This
asymmetry also appears to a much lesser degree in Figure 11 of
\citet{bec09}. Note that the animation that accompanies the 2-D simulation around a rapidly
spinning Kerr black hole in \citet{bec08} also shows asymmetric hairpin accretion. This supports
the notion that asymmetric hairpin field line accretion is not dependent on spin.
This does not mean that reconnection will proceed similarly in 2-D and 3-D and independent of spin.
Instead the simulations show only that the potential pre-reconnection field configuration
of asymmetric accretion of hairpin field lines seems to occur
in either 2-D or 3-D and for high spin and low spin.

\item Some of the hairpins can actually be buoyant near the horizon and move away from the black hole,
never penetrating the horizon (eg., the hairpin $\approx 30^{\circ}$
below the equator in Figure 11 of \citet{bec09} moves outward
between t= 14600 M and t= 14640 M)
\end{enumerate}

\section{Reconnection Geometry in the 3-D Accretion Near Rapidly Rotating Black Holes}
The reconnection aspect of the "coronal mechanism" described in
\citet{bec09} relies on the topology of poloidal loops in the inner
accretion disk. However, in the case of interest here, 3-D around a
rapidly spinning black hole, almost all the flux in the inner
accretion flow is twisted up into toroidal coils
\citep{hir04,pun10}. There are no simple poloidal loops in 3-D like
there were in the azimuthally averaged data in Figure 11 of
\citet{bec09}. Since topology is critical to reconnection, the 2-D
expedience is not implemented here for the sake of accuracy and the
expense of complexity. The analog of a 2-D poloidal loop in the
Schwarzschild geometry in the 3-D Kerr (rotating black hole)
geometry would seem to be one of the many twisted coils that
permeate the accretion flow \citet{hir04,pun10}. The "loop" from t =
9840 M (in geometrized units) of KDJ that is plotted in white in the
left hand frame of Figure 2 is typical of most of the field lines in
the inner accretion disk for high spin black holes. It spirals near
the black hole then as it expands vertically, it leaves the disk and
penetrates the corona. The twisted loop connects to large distances
through the corona, presumably closing far away. There are
variations of this topology in which one end of the coil stays in
the disk while the other end permeates the corona or event horizon
\citep{pun10}.
\begin{figure*}
\includegraphics[scale=0.19]{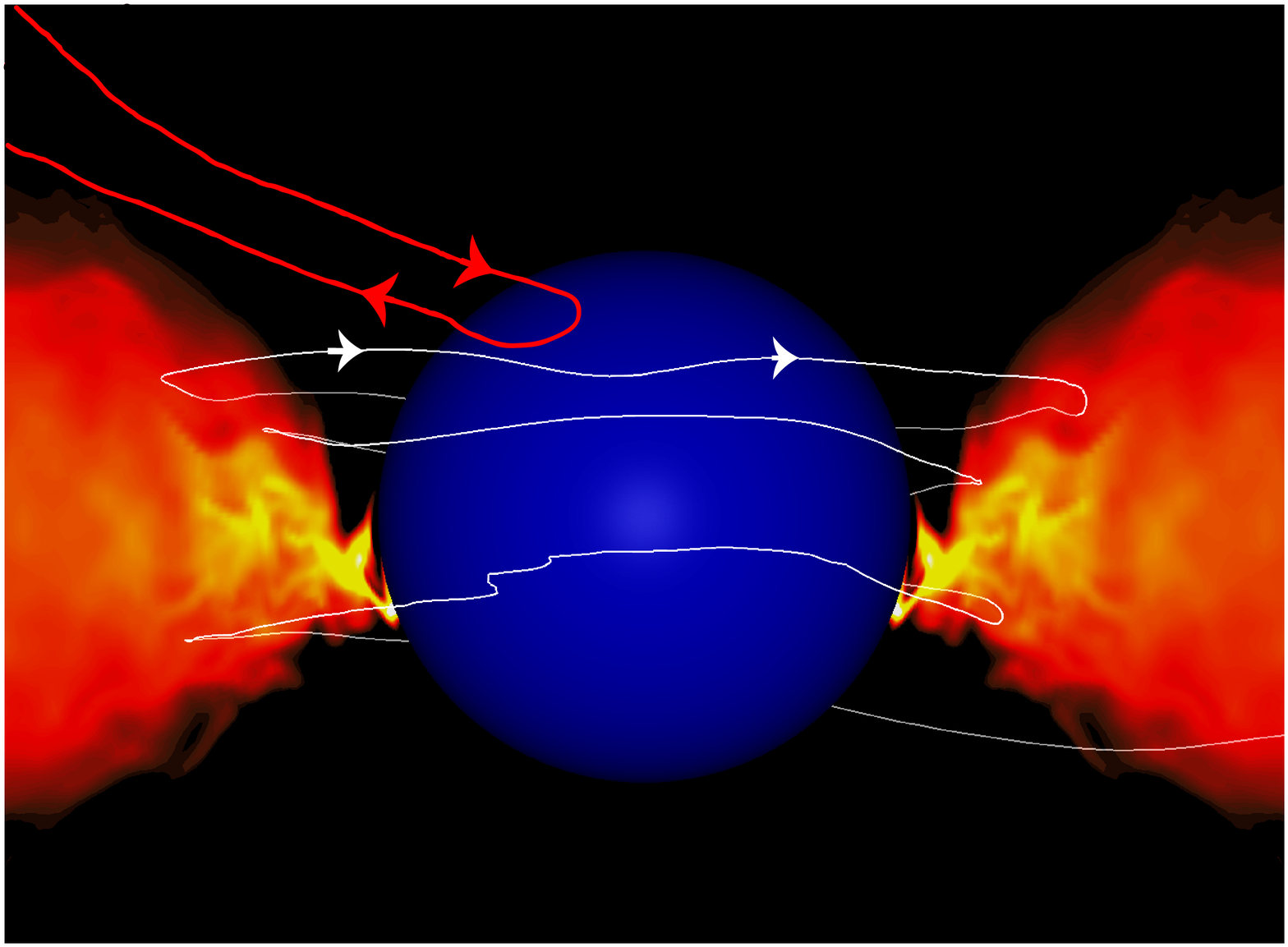}
\includegraphics[scale=0.19]{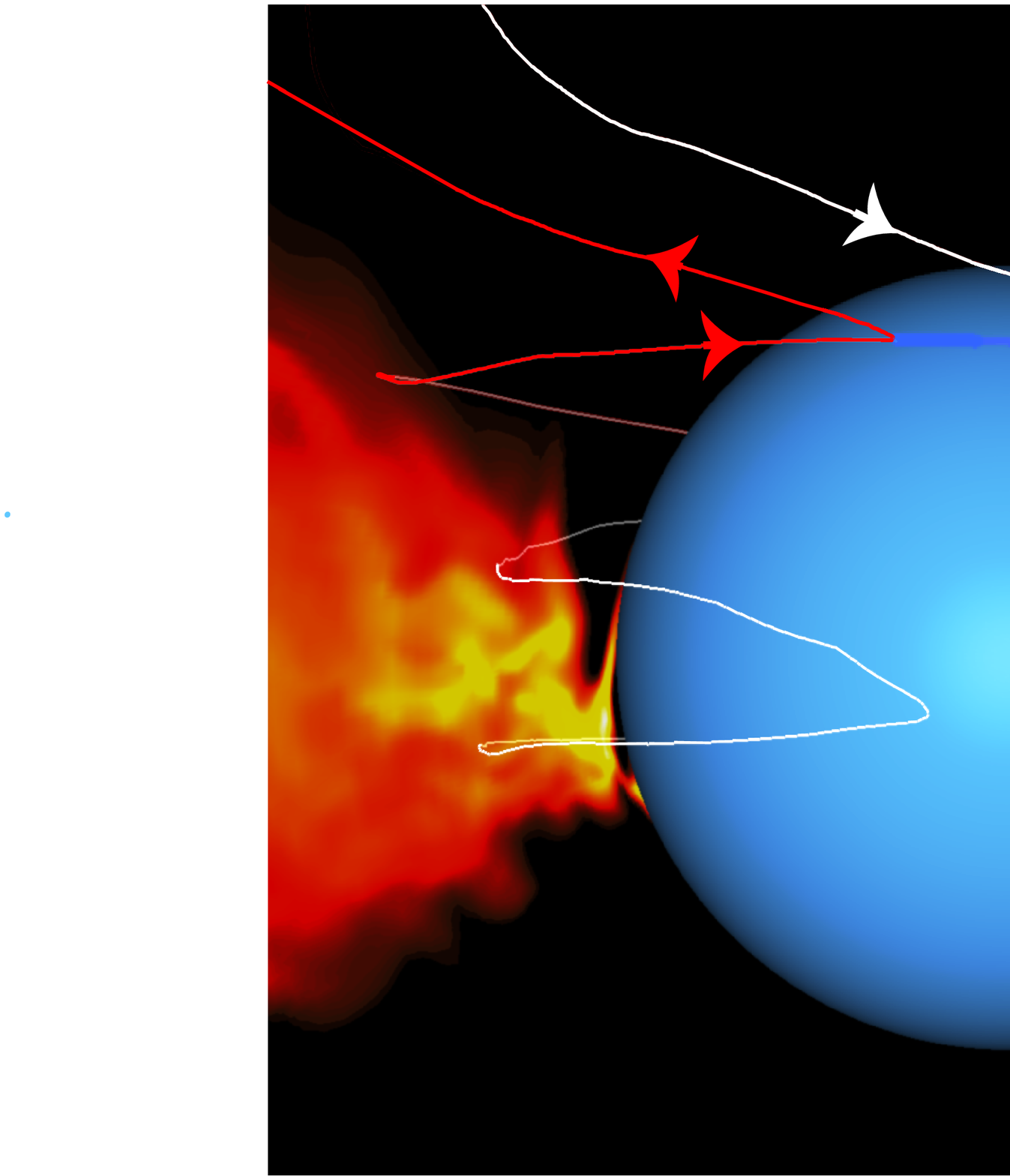}
\caption{A potential reconnection site in KDJ is displayed in the
left hand frame. The red field line is the accreting, slightly
twisted coronal hairpin. The white field line is a typical twisted
structure that is the 3-D equivalent of a poloidal loop in a 2-D
Schwarzschild representation. The inner calculational boundary is
blue. The post reconnection topology is depicted in the right hand
frame. The red curve is a buoyant hairpin. The white curve is a type
I field line that threads the ergospheric disk as in Figure 1.}
\end{figure*}
\par There are no fine time resolution snapshots of the simulation KDJ (the data is sampled every 80 M), so we cannot see
the asymmetric type I field lines of Figure 1 forming. There is only
circumstantial evidence as to the chain of events:
\begin{enumerate}
\item Most of the disk field lines in KDJ are twisted coils near the event horizon that can
have significant random inflections due to turbulence and never leave the disk or corona \citep{pun10,hir04}.
\item In KDJ, large vertical flux bundles are entrapped in the inner regions of the accretion flow tending
to be far more pronounced in one hemisphere \citep{pun10}.
\item In KDJ, the ergospheric disk flux shows no tendency for a preferred hemisphere. They
can be in either hemisphere or both hemispheres in an individual time snapshot \citep{pun10}.
\item The animations of simulations show that significant poloidal flux
can accrete to the inner regions of the accretion flow through the
corona in the form of hairpins in an episodic fashion \citep{bec08,bec09}.
\item These hairpins can arrive at the inner edge of the flow in asymmetric north/south configurations \citep{bec08,bec09}.
\item The topology of the poloidal flux near the event horizon was shown to be determined by reconnection
with simple poloidal loops in 2-D azimuthally averaged data \citep{bec09}.
\item Figure 11 of \citet{bec09} not only shows the asymmetric accretion of hairpins to the event horizon, but also
that some hairpins near the black hole can actually become buoyant and move away from the black hole.
\end{enumerate}
These seven facts can be used to consider the time evolution in the
twisted geometry of KDJ that results in the type I field line
topology in Figure 1. Figure 2 simulates the most plausible scenario
based on these seven results above which is used as a surrogate for
actual fine time scale data sampling. The left hand frame shows a
typical twisted accretion disk coil in white being approached by a
coronal hairpin in red near the black hole. Notice that the coronal
hairpin is azimuthally twisted in KDJ. Reconnection is very
complicated in a twisted 3-D environment and is not well understood
\citep{pon11}. However, the configuration as drawn forms a natural
reconnection site (an X-point). We expect both types of field lines
in Figure 2 to exist from points 1, 4 and 5 above. The elements
required for the pre-reconnection geometry in the left hand frame of
Figure 2, commonly occur in this family of simulations. Thus, it is
reasonable to expect that these field configurations coexist in
proximity at various times and these potential reconnection sites
should not be rare. However, the reconnection rate in such a
complicated geometry that does not proceed by a physical mechanism,
but through numerical diffusion, is very uncertain. By points 2, 3
and 6 above, reconnection must have occurred in KDJ as depicted in
the right hand frame of Figure 2. The white curve in the right hand
frame is poloidal flux through the equatorial plane of the
ergosphere in one hemisphere in analogy to the left hand frame of
Figure 1 and the red curve would be a buoyant hairpin field line
that moves out in the corona consistent with point 7.
\par A significant difference between the topology resulting from the reconnection in Figure 2 compared to that in Figure 11 of \citet{bec09},
is that in Figure 2 the reconnection is happening before the hairpin
penetrates the event horizon and in Figure 11 of \citet{bec09} it
occurs after the hairpin penetrates the horizon. This indicates that
the coherent flux transport rate combined with the reconnection rate
and twisted 3-D field line geometry (which affects the reconnection
rate) might determine if a field line penetrates the event horizon
or the inner accretion flow when and if reconnection occurs. The
final field line topology depends on the balancing of reaction rates
(reconnection and transport) as well as internal dynamics (that
affect field line shape) that are determined by the numerical
simulation.
\section{Discussion} The Letter shows that the "coronal mechanism" for flux transport in simulations
of black hole accretion provides a plausible explanation for the one sided
ergospheric disk field lines in the high spin 3-D simulation KDJ. It therefore explains the strange
phenomenon observed in KDJ that the black hole driven jet Poynting
flux was very one sided, jumping from side to side and emanating
primarily from the ergospheric disk.
\par An otherwise almost identical simulation to KDJ
that includes additional artificial diffusion terms in the equations
of continuity, energy conservation, and momentum conservation (as
described in \citet{dev07}) do not show these one sided ergospheric
disk structures \citet{pun11}. This is seems to indicate a change in
the the reconnection process that is driven either directly or
indirectly by the numerical diffusion. In support of this
interpretation, the force-free simulations of an initially uniform
field in \citet{kom04} show magnetic flux threading the ergospheric
equatorial plane near the black hole, yet the same initial state
that is time evolved in a different force-free code with a slower
ansatz for the reconnection rate shows no magnetic flux threading
the equatorial plane near the black hole \citep{mck05}. The
implication is that the global topology of the black hole
magnetosphere is highly dependent on the time evolution driven by
reconnection. This is not a trivial circumstance because in MHD
simulations reconnection is dependent on numerical diffusion. The
situation is rendered even more ambiguous by the complicated twisted
magnetic field line geometries in 3-D simulations around rapidly
rotating black holes (eg. Figure 2). The complications of far less
intricate 3-D field line topology have been recognized in solar and
planetary physics \citep{pon11,wil10}. Similarly, some detailed
numerical modeling has shown the need for 3-D to properly describe
reconnection \citep{kow09,kul10}. Furthermore, in high energy
environments radiation effects might also be crucial for a proper
treatment \citet{uzd11}. This might be relevant because radiation is
not formally considered in these MHD codes. In summary, there are
many potential sources of uncertainty in the reconnection induced
topology in these simulations.
\par It should be noted that in principle, the choice of coordinates
(eg. Boyer-Lindquist, as in the simulations discussed here, or
Kerr-Schild) should be irrelevant to the results obtained. What is
likely more relevant is the grid scale and sources of numerical
diffusion in the solution methodology. A very basic consideration is
that once a field line that is frozen-in to the plasma is accreted
through the event horizon (technically, the inner calculational boundary that is just
outside the event horizon in Boyer-Lindquist coordinates), it should be unable to migrate out of the
horizon by moving outward into the equatorial plane if causality is
maintained by the frozen-in plasma inside the horizon. Field lines can only be extracted from the horizon by local reconnection, not diffusion back into the equatorial plane. As such,
a black hole saturated with magnetic flux is a nontrivial "boundary
condition" for further magnetic flux accretion of the same
orientation. For example what happens when there is repeated injection
of magnetic flux as in the simulations of \citet{igu08}. The suite
of simulations considered here have a finite amount of flux (either vertical or the leading edge of accreting loops) that
accretes towards the black hole and a boundary condition that flux
cannot leave or enter through the outer boundary. However, if that
boundary condition is changed to be one of continuous injection of
flux then the magnetic topology might change near the black hole. In
particular, \citet{bec09} suggest that the funnel and horizon field strength is
set by the gas and magnetic pressure in the disk and \citet{tch10}
claim that it is set by the ram pressure in the disk. Once this very finite
value is achieved, what happens when more flux is injected into the
accretion stream and approaches the horizon? Does this preferentially move the reconnection
sites out to near the mid plane of the ergosphere? This might raise the power level to be
consistent with the more powerful radio loud AGN \citep{nem07,pun11}.

\par In the absence of more robust treatments of reconnection, this study
suggests two new simulations that are achievable with the present
numerical codes. One would be to rerun KDJ with the original code,
but with high density time sampling for a portion of the run so that
we can see how the ergospheric disk flux gets established. It would
also be an interesting simulation to explore the possibility of
continuous vertical flux injection from the outer boundary around
rapidly rotating black holes to see if the pattern of reconnection
changes and if the magnetospheric distribution of flux changes,
possibly enhancing the ergospheric disk. In the future, more
realistic reconnection in the presence of resistivity needs to
modeled near black holes as in \citet{pal09}. One might also find
environments suitable for reconnection in the ergospheres of other
simulation geometries that can produce relativistic jets. For
example, the collisions of black holes as discussed in \citet{pal10}
or the inner regions of a tilted accretion disk as modeled in
\citet{fra07} are fertile areas to pursue.

\end{document}